\documentclass[twocolumn,pra,showpacs,superscriptaddress,showpacs]{revtex4-1}
\usepackage{graphicx}
\usepackage{amssymb,amsmath}
\usepackage{bm}
\usepackage{dcolumn}
\usepackage{subfigure}
\usepackage{float}
\usepackage{url}
\usepackage{color}
\usepackage{txfonts}
\usepackage[dvipdfmx]{hyperref}
\hypersetup{backref,pdfpagemode=FullScreen,colorlinks=true,breaklinks,urlcolor=blue,linkcolor=blue,citecolor=blue}

\begin{document}
\title{Phase vortices of the quenched Haldane Model}

\author{Jinlong Yu}
\affiliation{Institute for Advanced Study, Tsinghua University, Beijing 100084, China}
\date{\today}

\begin{abstract}
Using the recently developed Bloch-state tomography technique, the quasimomentum $\bf k$-dependent Bloch states ${\left( {\sin \left( {{\theta _{\mathbf{k}}}/2} \right),\; - \cos \left( {{\theta _{\mathbf{k}}}/2} \right){e^{i{\phi _{\mathbf{k}}}}}} \right)^T}$ of a two-band tight-binding model with two sublattices can be mapped out. We show that, if we prepare the initial Bloch state as the lower-band eigenstate of a topologically trivial Haldane Hamiltonian $H_i$, and then quench the Haldane Hamiltonian to $H_f$, the time-dependent azimuthal phase ${\phi _{\mathbf{k}}(t)}$ supports two types of vortices. The first type of vortices are static, with the corresponding Bloch vectors pointing to the north pole ($\theta_{\mathbf{k}}=0$). The second type of vortices are dynamical, with the corresponding Bloch vectors pointing to the south pole ($\theta_{\mathbf{k}}=\pi$). In the $(k_x,k_y,t)$ space, the linking number between the trajectories of these two types of vortices equals exactly to the Chern number of the lower band of $H_f$, which provides an alternative method to directly map out the topological phase boundaries of the Haldane model.
\end{abstract}

\maketitle

\section{Introduction}
Topological insulators have attracted a lot of interests due to their novel properties and potential applications~\cite{Hasan2010, Qi2011}. One of the prototype of topological insulators is the Haldane model of Chern insulators~\cite{Haldane1988}, which describes spinless fermions hopping on a 2D honeycomb lattice, within a staggered magnetic field. With the shaking-lattice technique~\cite{Struck_Sengstock2011, Hauke_Lewenstein2012, Struck_Sengstock2012, Struck2013, Zheng2014}, the Haldane model has been realized directly in a recent cold atom experiment~\cite{Jotzu2014}, and the topological phase diagram is investigated using the Bloch oscillation method. It is worth mentioning that, although the Haldane model is realized in the cold atom setup, the topological ground state for this system is generally hard, if not impossible, to be reached. As the system is generally prepared in a topologically trivial state before the shaking is applied, by driving the system crossing the phase boundary to the topological phase, excitations are inevitable as the band gap closes at the phase boundary and adiabaticity could not be maintained.

The idea of a quantum quench, which starts with the ground state of a topologically trivial Hamiltonian $H_i$ supplemented by the evolution of a topologically trivial/nontrivial Hamiltonian $H_f$, provides an alterative way to studying the properties of the Haldane model (and Haldane-like topological models)~\cite{Caio-Cooper2015, Hu-Zoller2016, Mueller2016, Wilson2016, Pengfei2016}. Although the Chern number~\cite{TKNN1982} of the time-dependent state remains unchanged as the state evolution governed by $H_f$ is unitary~\cite{Rigol2015, Caio-Cooper2015}, the edge states~\cite{Caio-Cooper2015} as well as nontrivial Hall response~\cite{Hu-Zoller2016, Mueller2016, Wilson2016} can be built up dynamically.

It is interesting to see that, quantum quenches can also be used to perform Bloch-state tomography, as proposed in Ref.~\cite{Hauke_Lewenstein2014}, and then realized experimentally for a Haldane-like model on the honeycomb lattice by the Hamburg group~\cite{Sengstock2016a}. In this experiment, they prepared the initial (topologically trivial) state adiabatically, and then measured the Bloch state for each quasimomentum $\bf{k}$ parameterized on the Bloch sphere as ${\left( {\sin \left( {{\theta _{\mathbf{k}}}/2} \right),\; - \cos \left( {{\theta _{\mathbf{k}}}/2} \right){e^{i{\phi _{\mathbf{k}}}}}} \right)^T}$. The Berry curvature of the corresponding topologically trivial state is also mapped out. More recently, generalizing their previous scheme in Ref.~\cite{Sengstock2016a}, the Hamburg group performed a double-quench experiment and observed emergent phase vortices in the azimuthal phase $\phi_{\bf k}$~\cite{Sengstock2016}. To be more specific, they prepare the initial state as all the ${\bf k}$-dependent Bloch vectors pointing to the north pole regime $\left| {{\psi _{\mathbf{k}}}\left( {t = 0} \right)} \right\rangle  = {\left( {0,\;1} \right)^T}$ (i.e., $\theta_{\bf k}\to0$), then quench the Hamiltonian to $H_f$ giving rise to the time-dependent Bloch state $\left| {{\psi _{\mathbf{k}}}\left( t \right)} \right\rangle  = {e^{ - i{H_f}t}}\left| {{\psi _{\mathbf{k}}}\left( {t = 0} \right)} \right\rangle $. A second quench to a flat band Hamiltonian is performed to measure the time- and quasimomentum- dependent state parameterized by ${\theta _{\mathbf{k}}}(t)$, and ${\phi _{\mathbf{k}}}(t)$. They find that, two different types of vortices emerge in the phase profile of ${\phi _{\mathbf{k}}}(t)$. The first type is static vortices, which indicate the Dirac points of $H_f$. The locations of this kind of vortices do not move as the name \emph{static} implies. The second type is dynamical vortices. This kind of vortices emerge for particular quasimomentum points at particular times for some particular parameters of $H_f$. This kind of vortices can be pairwise created at some $\bf k$ points. They then move along some trajectories, and finally pairwise annihilate each other, making the trajectories closed. The appearance of dynamical vortices is identified as the dynamical order parameter for a dynamical phase transition~\cite{Heyl_DQPT_2013}, which  describes the singularities of the Loschmidt amplitude of the time-dependent state at a critical time. In addition to the identification of phase vortices, the dynamical phase diagram, which is defined as the parameter regime of the final Hamiltonian $H_f$ that support the appearance of dynamical vortices, is also explored. They find that, the region of the dynamical phase diagram is larger than the one of the equilibrium phase diagram of $H_f$. The dynamics of the phase vortices and the phenomena of enlarged dynamical phase diagram are not quantitatively discussed in Ref.~\cite{Sengstock2016}.

It is the purpose of the current work to explore the dynamics of the phase vortices for a quenched topological model. As the effective Hamiltonian (of a Haldane-like model) in Ref.~\cite{Sengstock2016} forbids a simple analytical expression, we study instead the phase vortices of the quenched Haldane model. We give explicitly the expression of the trajectories of the dynamical vortices, and give the possible reason for the enlarged dynamical phase diagram. We will show that, the trajectories of the dynamical vortices tell exactly the Chern number of the Hamiltonian $H_f$: when the dynamical vortices wind around static vortices, the lower-band Chern number of $H_f$ is $C_1 = 1$; while when the dynamical vortices do not wind around static vortices, we have $C_1 = 0$. More generally, as is proved in detail in Ref.~\cite{Pengfei2016}, the linking number between these two kinds of trajectories in the $(k_x, k_y, t)$ space equals to the Chern number of the lower band. Thus the quench protocol in Ref.~\cite{Sengstock2016} can be used to directly identify the topological phase boundaries of the equilibrium Hamiltonian.

The paper is organized as follows. In Sec.~\ref{Sec:2}, we describe the quench protocol of the Haldane model, and study the state evolution of this system for the case of infinite large initial energy offset $M_i\to\infty$ in the initial Hamiltonian $H_i$ (i.e., $\theta_{\bf k} = 0$). In Sec.~\ref{Sec:3}, the evolution of the time-dependent azimuthal phase ${\phi _{\mathbf{k}}}(t)$ is explored. We show that, once the south pole ($\theta_{\bf k} = \pi$) is reached, the dynamical vortices are pairwise created. The trajectories of the dynamical vortices as well as the dynamical phase diagram are identified. In Sec.~\ref{Sec:Finite_Mi}, we study the case of a finite initial energy offset $M_i$, and the enlarged dynamical phase diagram is achieved. We conclude in Sec.~\ref{Sec:Conclusion}.

\section{The state evolution of the quenched Haldane model}\label{Sec:2}
\begin{figure}[b]
  \includegraphics[width=\columnwidth]{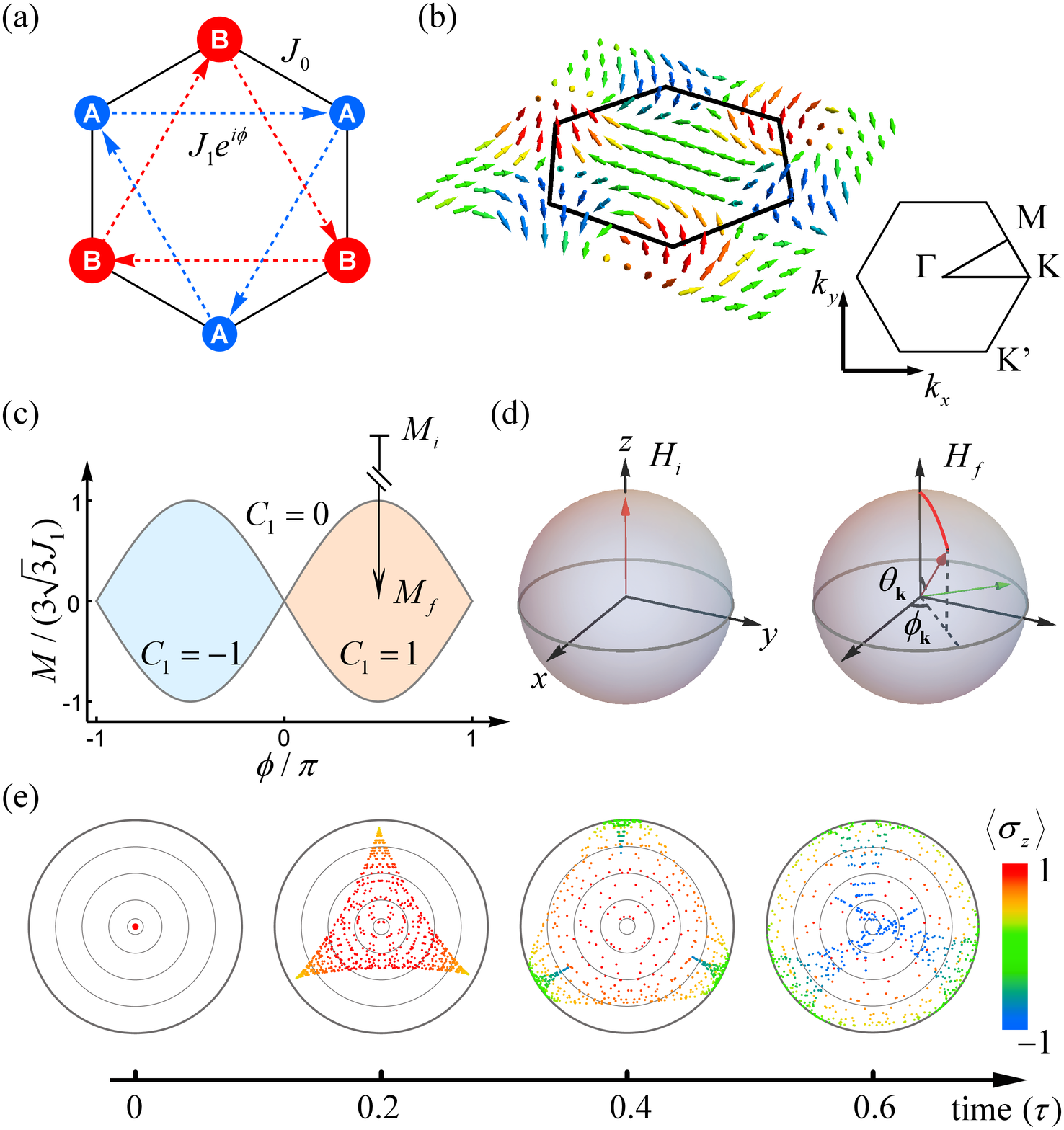}\\
  \caption{Quench dynamics of the Haldane model. (a) The Haldane model on a honeycomb lattice. The two sublattices $A$ and $B$ take an energy offset $M$. The arrows indicate the tunneling coefficients with positive hopping phases. (b) The normalized vector ${\mathbf{\hat h}}(\mathbf{k}) = {\mathbf{h}}/|{\mathbf{h}}|$ in the quasimomentum space for the topologically nontrivial case. The hexagon shows the first Brillouin zone. (c) The topological phase diagram of the Haldane model and the quench protocol. Sublattice energy offset $M$ is quenched from $M_i$ to $M_f$ for a particular $\phi$. (d) The state evolution visualized on the Bloch sphere for a particular quasimomentum point $\bf{k}$.  (e) The state evolution for quasimomentum points in the first Brillouin zone, viewed from the north pole. The south pole region is reached for time $t\sim0.6\tau$. Here we take $\phi = \pi/2$, $M_i \to \infty$, $M_f = 0.1 J_0$, $J_1 = 0.1 J_0$, and $\tau=1/J_0$. }\label{Fig1}
\end{figure}

We consider the Haldane model~\cite{Haldane1988} on the honeycomb lattice [Fig.~\ref{Fig1}(a)] with the following Hamiltonian:
\begin{equation}
  \begin{aligned}
  H =&  - {J_0}\sum\limits_{j,l} {\left( {a_{{{\mathbf{r}}_j}}^\dag {b_{{{\mathbf{r}}_j} + {{\bm{\delta }}_l}}} + {\text{h}}{\text{.c}}{\text{.}}} \right)}  + M\sum\limits_j {\left( {a_{{{\mathbf{r}}_j}}^\dag {a_{{{\mathbf{r}}_j}}} - b_{{{\mathbf{r}}_j}}^\dag {b_{{{\mathbf{r}}_j}}}} \right)}  \hfill \\
   &- {J_1}\sum\limits_{j,l} {\left( {{e^{i{\phi _{jl}}}}a_{{{\mathbf{r}}_j}}^\dag {a_{{{\mathbf{r}}_j} + {{\mathbf{a}}_l}}} + {e^{ - i{\phi _{jl}}}}b_{{{\mathbf{r}}_j}}^\dag {b_{{{\mathbf{r}}_j} + {{\mathbf{a}}_l}}}} \right)} , \hfill \\
\end{aligned}
\end{equation}
where $a_{{{\mathbf{r}}_j}}$ and $b_{{{\mathbf{r}}_j}}$ are the annihilation operators for the $A$ and $B$ sublattices with coordinates ${{\mathbf{r}}_j}$ respectively, $J_0$ and $J_1$ are respectively the nearest-neighboring (NN) and next-nearest-neighboring (NNN) hopping amplitude, $\bm{\delta}_l$ ($\mathbf{a}_l$) with $l=1,2,3$ are the vectors pointing to the NN (NNN) sites, $M$ is the sublattice energy offset. Here $\phi_{jl}=\pm\phi$ are the NNN hopping phases, and the positive sign is taken for the NNN hopping along the arrows as shown in Fig.~\ref{Fig1}(a). We set $|\bm{\delta}_l| = 1$ as the length unit.
The Fourier transformed Hamiltonian in the quasimomentum space
gives the following compact form
\begin{equation}
  H = \sum\limits_{\mathbf{k}} {\left( {a_{\mathbf{k}}^\dag \quad b_{\mathbf{k}}^\dag } \right)} \mathcal{H}\left( {\mathbf{k}} \right)\left( \begin{gathered}
  {a_{\mathbf{k}}} \hfill \\
  {b_{\mathbf{k}}} \hfill \\
\end{gathered}  \right),
\end{equation}
where ${\mathbf{k}}$ is the quasimomentum, $a_{\mathbf{k}}$ and $b_{\mathbf{k}}$ are the sublattice annihilation operators.
In the above expression, $\mathcal{H}\left( {\mathbf{k}} \right)$ is a $2\times2$ matrix with the form
\begin{equation}\label{H_eff}
  \mathcal{H}\left( {\mathbf{k}} \right) = {h_0}\left( {\mathbf{k}} \right){I_2} + {\mathbf{h}}\left( {\mathbf{k}} \right) \cdot {\bm{\sigma}},
\end{equation}
where $I_2$ is the $2\times2$ unit matrix, ${\bm{\sigma}}$ is the vector of Pauli matrices, ${h_0} =  - 2{J_1}\cos \phi \sum\nolimits_l {\cos \left( {{\mathbf{k}} \cdot {{\mathbf{a}}_l}} \right)}$, and ${\mathbf{h}}\left( {\mathbf{k}} \right) = (h_x, h_y, h_z)$ with
\begin{equation}\label{h0_hxyz_haldane}
  \begin{gathered}
  {h_x} =  - {J_0}\sum\limits_l {\cos \left( {{\mathbf{k}} \cdot {{\bm{\delta }}_l}} \right)} ,\quad {h_y} =  - {J_0}\sum\limits_l {\sin \left( {{\mathbf{k}} \cdot {{\bm{\delta }}_l}} \right)} , \hfill \\
  {h_z} = M - 2{J_1}\sin \phi \sum\limits_l {\sin \left( {{\mathbf{k}} \cdot {{\mathbf{a}}_l}} \right)} . \hfill \\
\end{gathered}
\end{equation}
The spectrum of the system is ${E_{\mathbf{k}}} = {h_0} \pm h$, where $h=\sqrt {h_x^2 + h_y^2 + h_z^2} $. The topology of the system can be characterized by the (first) Chern number~\cite{TKNN1982} of the lowest band as follows~\cite{Qi2006}
\begin{equation}
  {C_1} = \iint_{{\text{BZ}}} {\frac{{{d^2}k}}{{4\pi }}}{\mathbf{\hat h}} \cdot \left( {\frac{{\partial {\mathbf{\hat h}}}}{{\partial {k_x}}} \times \frac{{\partial {\mathbf{\hat h}}}}{{\partial {k_y}}}} \right),
\end{equation}
where ${\mathbf{\hat h}} = {\mathbf{h}}/h$. The nontrivial winding of ${\mathbf{\hat h}}$ [see Fig.~\ref{Fig1}(b)] exists in the parameter regime $|M| < 3\sqrt 3 |{J_1}\sin \phi |$, which gives the phase diagram of the system [Fig.~\ref{Fig1}(c)].

We now consider the case that the sublattice energy offset $M$ is time dependent:
\begin{equation}
  M(t) = \left\{ \begin{gathered}
  {M_i},\quad t \leqslant 0; \hfill \\
  {M_f},\quad t > 0. \hfill \\
\end{gathered}  \right.
\end{equation}
Suppose the system is initially prepared in the ground state of the Haldane Hamiltonian $H_i=h_0{I_2}+{\bf{h}}_i\cdot\bm{\sigma}$ with offset $M_i$, after the quench at $t=0$ the state evolves under a new Hamiltonian $H_f=h_0{I_2}+{\bf{h}}_f\cdot\bm{\sigma}$ with offset $M_f$ [Fig.~\ref{Fig1}(d)]. We then want to ask the following question: could the time-dependent state give any insight of the topological property of the system? The answer is \emph{Yes}. The  time-dependent state, albeit always is topologically trivial, can indeed be used to identify the topological phase diagram of the equilibrium Hamiltonian~\cite{Pengfei2016}.

\begin{figure*}
  \includegraphics[width=\textwidth]{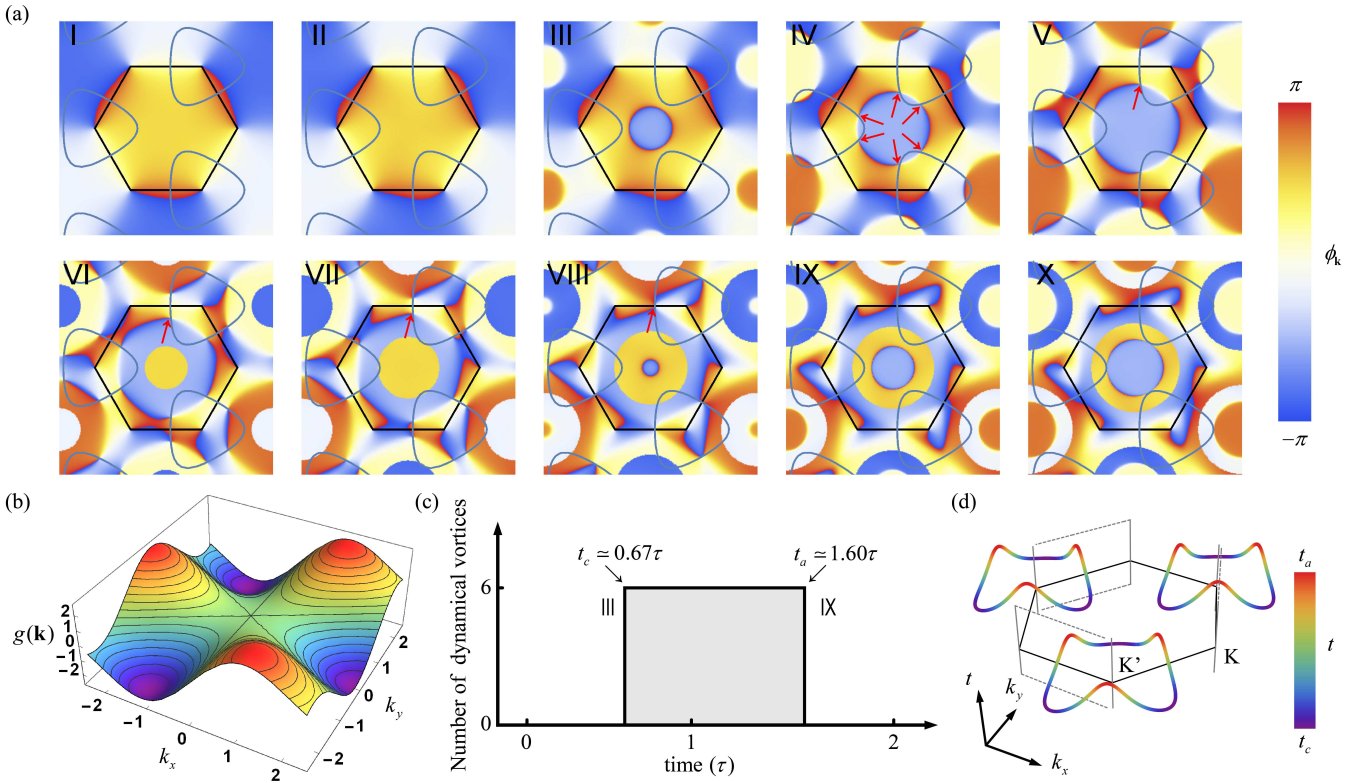}\\
  \caption{Dynamics of the azimuthal phase $\phi_{\bf{k}}(t)$, with parameters the same as in Fig.~\ref{Fig1}(e). (a) The stroboscopic observation of $\phi_{\bf k}(t)$, for $t = 0.2\tau$ (I), $0.4\tau$ (II), $...$, $2\tau$ (X). The hexagon shows the first Brillouin zone. The closed loops show the contours with $h_z({\bf k}) = 0$. Dynamical vortices, as indicated by the red arrows in (IV, V, VI, VII and VIII), live on such contours within particular time slots. (b) The $g({\bf{k}})$ function [Eq.~(\ref{g_k_function})] in the quasimomentum space. $g({\bf{k}})=0.5=M_f/(2J_1 \sin\phi)$ gives the nodal lines of $h_z({\bf k})$ as shown in (a). (c) The time dependent dynamical vortex number. (d) The vortex trajectories in the $(k_x,k_y,t)$ space. The linking number between the loop of dynamical vortices and the loop of static vortices equals to the Chern number of the final Hamiltonian $C_1=1$ for this case. }\label{Fig2}
\end{figure*}

For conceptual simplicity, we first consider $M_i\to+\infty$, and defer the more complex case of finite $M_i$ to Sec.~\ref{Sec:Finite_Mi}. For the $M_i\to+\infty$ case, the initial single-particle state for each quasimomentum point reads
\begin{equation}
  \left| {{\psi _{\mathbf{k}}}\left( {t = 0} \right)} \right\rangle  = \left( \begin{gathered}
  0 \hfill \\
  1 \hfill \\
\end{gathered}  \right).
\end{equation}
Under the Hamiltonian ${H}_f\left( {\mathbf{k}} \right) = {h_0}\left( {\mathbf{k}} \right){I_2} + {\mathbf{h}_f}\left( {\mathbf{k}} \right) \cdot {\bm{\sigma}}$, we get (we always take the reduced Planck constant $\hbar=1$)
\begin{equation}\label{Eq:psi_t}
  \begin{aligned}
  \left| {{\psi _{\mathbf{k}}}\left( t \right)} \right\rangle  &= {e^{ - iH_f\left( {\mathbf{k}} \right)t}}\left| {{\psi _{\mathbf{k}}}\left( {t = 0} \right)} \right\rangle  \hfill \\
   &= {e^{ - i{h_0}t}}\left( \begin{gathered}
   - i\sin \left( {ht} \right)\left( {{{\hat h}_x} - i{{\hat h}_y}} \right) \hfill \\
  \cos \left( {ht} \right) + i\sin \left( {ht} \right){  {{\hat h}_z}} \hfill \\
\end{gathered}  \right). \hfill \\
\end{aligned}
\end{equation}
In the above, as the initial state is $\mathbf{k}$-independent, the subscript $f$ is omitted to keep notation concise.
This state can be parameterized on a Bloch sphere as
\begin{equation}\label{Eq:psi_theta_phi}
  \left| {{\psi _{\mathbf{k}}}\left( t \right)} \right\rangle : = \left( \begin{gathered}
  \sin \left( {{\theta _{\mathbf{k}}}/2} \right) \hfill \\
   - \cos \left( {{\theta _{\mathbf{k}}}/2} \right){e^{i{\phi _{\mathbf{k}}}}} \hfill \\
\end{gathered}  \right),
\end{equation}
then we have
\begin{equation}\label{Eq:phi_k}
  \left\{ \begin{gathered}
  {\theta _{\mathbf{k}}} = \arccos \left( {\cos {{\left( {ht} \right)}^2} + \sin {{\left( {ht} \right)}^2}\left( {\hat h_z^2 - \hat h_x^2 - \hat h_y^2} \right)} \right), \hfill \\
  {\phi _{\mathbf{k}}} = \arg \left[ {\frac{{\cos \left( {ht} \right) + i\sin \left( {ht} \right){{\hat h}_z}}}{{i\sin \left( {ht} \right)\left( {{{\hat h}_x} - i{{\hat h}_y}} \right)}}} \right]. \hfill \\
\end{gathered}  \right.
\end{equation}
The time- and quasimomentum-dependent Bloch vectors are then given by
\begin{equation}
  {{\mathbf{n}}_{\mathbf{k}}}\left( t \right) = \left( {\sin {\theta _{\mathbf{k}}}\cos {\phi _{\mathbf{k}}},\sin {\theta _{\mathbf{k}}}\sin {\phi _{\mathbf{k}}},\cos {\theta _{\mathbf{k}}}} \right).
\end{equation}
The typical results for the state evolution visualized by the Bloch vectors are given in Fig.~\ref{Fig1}(e).
As indicated in Fig.~\ref{Fig1}(e), some of the Bloch vectors can reach the south pole region as time goes by. As the effective magnetic field [${\mathbf{h}}\left( {\mathbf{k}} \right)$ in Eq.~(\ref{H_eff})] that drives the rotation of Bloch vector is orthogonal to the latter, it should have a vanishing $z$ component for this particular case, i.e., we should have $h_z({\bf k}) = 0$. According to Eq.~(\ref{h0_hxyz_haldane}), we rewrite $h_z$ as
  ${h_z({\bf k})} = M_f - 2{J_1}\sin(\phi) g({\bf k})$,
where $g({\bf k})$ is given by
\begin{equation}\label{g_k_function}
  g({\bf k}) = {\sin \left( {{\mathbf{k}} \cdot {{\mathbf{a}}_1}} \right) + \sin \left( {{\mathbf{k}} \cdot {{\mathbf{a}}_2}} \right) + \sin \left( {{\mathbf{k}} \cdot {{\mathbf{a}}_3}} \right)}.
\end{equation}
We plot $g({\bf k})$ in Fig.~\ref{Fig2}(b). We see that, $g({\bf k})$ peaks at $\pm K$ points, and thus is bounded as $|g({\bf k})|\le |g(k_x = \frac{4\pi}{3\sqrt{3}}, k_y = 0)| = \frac{3\sqrt{3}}{2}$. We thus see that, only when the Hamiltonian is in the topological phase (i.e., $|M_f| < 3\sqrt 3 |{J_1}\sin \phi |$) does the $h_z({\bf k})$ function features zeros.

\section{The evolution of the azimuthal phase: static and dynamical vortices}\label{Sec:3}
We now focus on the evolution of the time- and quasimomentum-dependent azimuthal phase $\phi_{\bf k}(t)$, as shown in Fig.~\ref{Fig2}(a). The main features of the azimuthal phase dynamics is as follows:
  (i) There are always static vortices in the phase pattern at the corners of the Brillouin zone (K and K' points) in Fig.~\ref{Fig2}(a).
  (ii) There are dynamical vortices that emerge [Fig.~\ref{Fig2}(a)IV], propagate [Fig.~\ref{Fig2}(a)(V-VI)], and annihilate [Fig.~\ref{Fig2}(a)VII] in the quasimomentum space.
  (iii) The trajectories of the dynamical vortices are just the zeros of $h_z({\bf k})$.

The locations of the static vortices correspond to the Dirac points of $h_x({\bf k})\sigma_x + h_y({\bf k})\sigma_y$, i.e., the $K$ and $K'$ points. Now that we have $h_x({\bf k}=K) = h_y({\bf k}=K) = 0$, the Bloch vectors of these points always point to the north pole.
We have shown that, only for the points satisfying $h_z({\bf k})=0$, the south pole is reachable at a suitable time. An accompanied question is that:
\begin{quote}
    \emph{For a particular quasimomentum point $\bf k$, must there be a dynamical vortex once the south pole is reached?}
\end{quote}
This question can be also reformulated quantitatively as follows:
Consider a particular quasimomentum point $\left( {k_x^0,k_y^0} \right)$ at a particular time $t_v$ that satisfy
\begin{equation}\label{Eq:condition_for_south_pole}
  \left\{ \begin{gathered}
  {h_z}\left( {k_x^0,k_y^0} \right) = 0, \hfill \\
  \cos \left( {h(k_x^0,k_y^0){t_v}} \right) = 0, \hfill \\
\end{gathered}  \right.
\end{equation}
what is the vorticity $\nu$ for this point? The answer is (see Appendix~\ref{App:South-pole} for details)
\begin{equation}\label{Eq:vorticity}
  \nu  = {\text{sign}}\left[ {\frac{{\partial \left( { h,{{\hat h}_z}} \right)}}{{\partial \left( {{k_x},{k_y}} \right)}}{|_{(k_x^0,k_y^0)}}} \right].
\end{equation}
Thus we see that, as long as the Jacobian ${\frac{{\partial \left( { h,{{\hat h}_z}} \right)}}{{\partial \left( {{k_x},{k_y}} \right)}}{|_{(k_x^0,k_y^0)}}}$ is nonvanishing, there is always a vortex at $(k_x^0,k_y^0)$ at time $t_v$. And we can also show (see Appendix~\ref{App:South-pole}) that the vortex dynamics features a time periodicity ${t_{v,n}} = \left( {1 + 2n} \right){t_{v,0}}$, where ${t_{v,0}} = \frac{1}{{h(k_x^0,k_y^0)}}\frac{\pi }{2}$ is the first time that the dynamical vortex emerge at the point $(k_x^0,k_y^0)$.

It is also checked that, at the particular points that the vortices emerge or annihilate, the corresponding Jacobians do vanish. For instance, we consider the system shown in Fig.~\ref{Fig2} [i.e., in Fig.~\ref{Fig1}(e)].  At the point $(k_x^0 \simeq  - 0.97,k_y^0 = 0)$,  we have $h_z(k_x^0 \simeq  - 0.97,k_y^0 = 0)=0$ and $\frac{{\partial h}}{{\partial {k_y}}}{|_{(k_x^0,k_y^0)}} = 0 = \frac{{\partial {{\hat h}_z}}}{{\partial {k_y}}}{|_{(k_x^0,k_y^0)}}$. This is also the point that the vortices are created at the time ${t_c} = \frac{1}{{h(k_x^0\simeq  - 0.97,k_y^0=9)}}\frac{\pi }{2} \simeq 0.67\tau$. Similarly, at the point $(k_x^0 \simeq 0.15,k_y^0 = 2\pi /3)$ (located at the edge of the Brillouin zone), we also have Eq.~(\ref{Eq:condition_for_south_pole}) satisfied with zero vorticity at the vortex annihilation time ${t_a} \simeq 1.60\tau$. Between time $t_c$ and $t_a$, we have three pairs of dynamical vortices as shown in Fig.~\ref{Fig2}(c).
It is also interesting to see the vortex trajectories in the $(k_x,k_y,t)$ space [Fig.~\ref{Fig2}(d)] which is related to the Hopf mapping and linking numbers~\cite{Pengfei2016}: the trajectories of dynamical vortices always wind around the Dirac points where the static vortices locate.

\begin{figure}
  \includegraphics[width=\columnwidth]{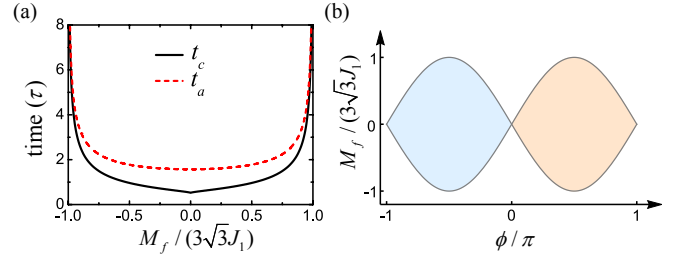}\\
  \caption{(a) The vortex creation (annihilation) time $t_c$ ($t_a$) as a function of $M_f$, for $\phi = \pi/2$, $M_i \to \infty$, and $J_1 = 0.1J_0$. They diverge at $M_f / (3\sqrt(3)J_1)=\pm1$, which indicates that for $|M_f / (3\sqrt{3}J_1)|\ge1$ case, no dynamical vortex can be created. (b) The dynamical phase diagram with the shaded regions hosting dynamical vortices for the $M_i \to \infty$ case.  }\label{Fig3}
\end{figure}

The arguments on the vortex creation and annihilation time for the example can be extended to the whole phase diagram of the system. For instance, we fix $\phi = \pi/2$ and $J_1 = 0.1J_0$, by varying the sublattice imbalance $M_f$, we can get the vortex creation time $t_c$ and annihilation time $t_a$ as a function of $M_f/J_1$, as shown in Fig.~\ref{Fig3}(a). The primary information in Fig.~\ref{Fig3}(a) is that, the dynamical vortices only appear when the evolution Hamiltonian is topologically nontrivial, as $h_z(\mathbf{k})$ does not possess nodal lines otherwise.
It is seen that, $t_c$ (and hence $t_a$) diverges as the parameters approach the phase boundary of the static phase diagram [$M/\left( {3\sqrt 3 {J_1}\sin \phi } \right) =  \pm 1$]. This statement is also true for a general $\phi$, and we thus get the {\lq\lq}dynamical phase diagram{\rq\rq} (defined as the parameter region that dynamical vortices can appear~\cite{Sengstock2016}) as shown in Fig.~\ref{Fig3}(b).  This dynamical phase diagram coincides exactly with the phase diagram of the equilibrium Hamiltonian [Fig.~\ref{Fig1}(c)].

\section{Finite initial offset $M_i$: enlarged dynamical phase diagram}\label{Sec:Finite_Mi}
We now briefly discuss the case of preparing the initial state as the Hamiltonian with a finite (positive) $M_i$ in the topologically trivial phase. In the Bloch sphere representation, the initial Bloch vectors now point to a finite region lies in the northern hemisphere. We can use ${\Theta _{\mathbf{k}}}$ and ${\Phi_{\mathbf{k}}}$ to parameterize the Hamiltonian vector ${\mathbf{\hat h}}\left( {\mathbf{k}} \right)$ as
\begin{equation}
  {\mathbf{\hat h}}\left( {\mathbf{k}} \right) = (\sin {\Theta _{\mathbf{k}}}\cos {\Phi _{\mathbf{k}}},\sin {\Theta _{\mathbf{k}}}\sin {\Phi _{\mathbf{k}}},\cos {\Theta _{\mathbf{k}}}).
\end{equation}
Then it can be shown (see Appendix~\ref{App:Finite_Mi}) that the above discussion of vortex dynamics with $M_i\to\infty$ in Sec.~\ref{Sec:3} and Appendix~\ref{App:South-pole} is still valid as long as we replace the vortex trajectory ${h_z}\left( {\mathbf{k}} \right) = 0$ by
\begin{equation}\label{Eq:hz_tilde}
  {\hat{\tilde {h}}_z}\left( {\mathbf{k}} \right) = \cos \left( {\Theta _{\mathbf{k}}^f - \frac{{\Theta _{\mathbf{k}}^i}}{2}} \right) = 0.
\end{equation}
It recovers the previous results for $M_i\to\infty$, as ${\Theta _{\mathbf{k}}^i}=0$ and ${\hat{\tilde h}_z}\to{{\hat h}_z}$ for this case.

\begin{figure}
  \includegraphics[width=\columnwidth]{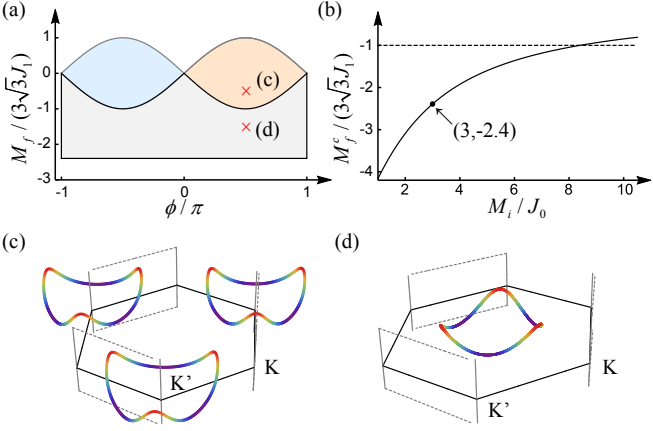}\\
  \caption{The vortex dynamics with a finite initial energy offset $M_i$. (a) Enlarged dynamical phase diagram for the $M_i=3J_0$ case. Apart from the colored parameter region shown in Fig.~\ref{Fig3}(b), there exists a new gray region which is bounded by a $\phi$-independent critical final energy offset $M_f^c$. For the $M_f^c<M_f<0$ case, a second type of dynamical vortices emerges. (b) The critical final energy offset $M_f^c$ for different initial offset $M_i$. (c, d) The vortex trajectories in the $(k_x,k_y,t)$ space with parameters marked by the red crosses in (a). They give rise to linking numbers $1$ and $0$ respectively. }\label{Fig4}
\end{figure}

For this finite $M_i$ case, the vortex dynamics is summarized in Fig.~\ref{Fig4}. The dynamical phase diagram is enlarged with an additional parameter region bounded by a $\phi$-independent critical $M_f^c$ [Fig.~\ref{Fig4}(a)]. Two types of vortex trajectories are possible as shown in Fig.~\ref{Fig4}(c) and (d), with different linking numbers. The trajectory of dynamical vortices as shown in Fig.~\ref{Fig4}(d) shrinks to a point at $\Gamma$ point at critical $M_f^c$, and the relation between $M_i$ and $M_f^c$ is given by
\begin{equation}
  \Theta _{{\mathbf{k}} = \Gamma }^f - \frac{{\Theta _{{\mathbf{k}} = \Gamma }^i}}{2} = \frac{\pi }{2},
\end{equation}
or, written explicitly,
\begin{equation}
  M_f^c =  - \tan \left[ {\frac{1}{2}\arccos \left( {\frac{{{M_i}}}{{\sqrt {9J_0^2 + M_i^2} }}} \right)} \right],
\end{equation}
as shown in Fig.~\ref{Fig4}(b). We see in Fig.~\ref{Fig4}(b) that, the critical value $M_f^c$ can be larger than $-3\sqrt{3}J_1$ as indicated by the dashed line, and it turns to zero for extremely large initial offset $M_i\to\infty$, which returns to the situation in Fig.~\ref{Fig3}(b). Indeed, the parameter regime $ M_f^c\le M_f \le 0$ just indicates the fact that the dynamical vortices around $\Gamma$ point can emerge for geometric reasons. Such type of dynamical vortices can also coexist with the ones winding around static vortices in the topological phase of the equilibrium Hamiltonian, as the former always give null contribution to linking numbers.

\section{Conclusion}\label{Sec:Conclusion}
Inspired by the recent Hamburg group experiment on the creation of dynamical vortices in a quenched Haldane-like model on a driven honeycomb lattice~\cite{Sengstock2016}, we have studied the dynamics of the phase vortices of the quenched Haldane model in detail. We have shown that, two types of vortices can appear in the azimuthal phase profile: the static vortices correspond to the Bloch states pointing to the north pole, while the dynamical vortices correspond to the Bloch states pointing to the south pole. The trajectories of the dynamical vortices also constitute of two types: one winds around Dirac point ($K$ or $K$'); the other winds around $\Gamma$ point, which could give rise to the enlarged dynamical phase diagram. The former gives linking number one, while the later gives linking number zero. As the linking number equals to the Chern number exactly as proved in Ref.~\cite{Pengfei2016}, the trajectories of the dynamical vortices provide an alternative method to determine the topological phase diagram of the Haldane model. The discussion in this paper can also be applied to the topological square-lattice model as realized recently by the USTC group~\cite{USTC_SOC2016}.
\\

\begin{acknowledgments}
The author would like to thank Xin Chen, Ce Wang, Hui Zhai, and Pengfei Zhang for valuable discussions.
\end{acknowledgments}

\appendix
\section{The vorticity for the point reaching the south pole}\label{App:South-pole}
In this appendix, we give the reduction of the vorticity for a particular quasimomentum point $\left( {k_x^0,k_y^0} \right)$ at the time $t_v$ when the south pole is reached. With the conditions stated in Eq.~(\ref{Eq:condition_for_south_pole}), the corresponding Bloch state, according to Eq.~(\ref{Eq:psi_t}), is given by (apart from an irrelevant overall phase)
\begin{equation}
  \left| {{\psi _{(k_x^0,k_y^0)}}\left( {t = {t_v}} \right)} \right\rangle  = \left( \begin{gathered}
   - i\left( {{{\hat h}_x} - i{{\hat h}_y}} \right) \hfill \\
  0 \hfill \\
\end{gathered}  \right).
\end{equation}
Such a state exactly points to the south pole. By solving the second equation in Eq.~(\ref{Eq:condition_for_south_pole}), we get \begin{equation}
  {t_{v,n}} = \frac{1}{{h(k_x^0,k_y^0)}}\frac{\pi }{2}\left( {1 + 2n} \right),
\end{equation}
where $n$ is a non-negative integer. We then consider, at the time $t_{v,n}$, the states nearby the quasimomentum point $\left( {k_x^0,k_y^0} \right)$. For simplicity, we take $n=0$, and thus set ${t_v} = t_{v,0} = \frac{1}{{h(k_x^0,k_y^0)}}\frac{\pi }{2}$. At the point $({k_x},{k_y}) = (k_x^0 + q\cos \alpha ,k_y^0 + q\sin \alpha )$, where $q$ is an infinitesimal real variable, we expand ${\hat h}_z({k_x},{k_y})$ and $\cos(h({k_x},{k_y})t_v)$ as follows:
\begin{equation}
  \begin{gathered}
  {{\hat h}_z}\left( {{k_x},{k_y}} \right) = q\left( {\frac{{\partial {{\hat h}_z}}}{{\partial {k_x}}}\cos \alpha  + \frac{{\partial {{\hat h}_z}}}{{\partial {k_y}}}\sin \alpha } \right){|_{(k_x^0,k_y^0)}}, \hfill \\
  \cos \left( {({k_x},{k_y}){t_v}} \right) =  - \frac{\pi }{2h}q\left( {\frac{{\partial  h}}{{\partial {k_x}}}\cos \alpha  + \frac{{\partial  h}}{{\partial {k_y}}}\sin \alpha } \right){|_{(k_x^0,k_y^0)}}, \hfill \\
\end{gathered}
\end{equation}
where the partial derivatives are evaluated at the point $(k_x^0,k_y^0)$. We thus have
\begin{equation}
\left| {{\psi _{({k_x},{k_y})}}\left( {t = {t_v}} \right)} \right\rangle  = \left( \begin{gathered}
   - i\left( {{{\hat h}_x} - i{{\hat h}_y}} \right) \hfill \\
  qf\left( \alpha  \right) \hfill \\
\end{gathered}  \right),
\end{equation}
where we have defined
\begin{equation}
  \begin{aligned}
  f\left( \alpha  \right) =  &- \frac{\pi }{{2h}}\left( {\frac{{\partial h}}{{\partial {k_x}}}\cos \alpha  + \frac{{\partial h}}{{\partial {k_y}}}\sin \alpha } \right){|_{(k_x^0,k_y^0)}} \hfill \\
  &  + i\left( {\frac{{\partial {{\hat h}_z}}}{{\partial {k_x}}}\cos \alpha  + \frac{{\partial {{\hat h}_z}}}{{\partial {k_y}}}\sin \alpha } \right){|_{(k_x^0,k_y^0)}}. \hfill \\
\end{aligned}
\end{equation}
Away from the $\pm K$ points, ${{{\hat h}_x} - i{{\hat h}_y}}$ is a slow-varying function of $q$, thus we have the azimuthal phase given by
\begin{equation}
  {\phi _{({k_x},{k_y})}}\left( {t = {t_v}} \right)\xrightarrow{{q \to 0}}\arg \left[ {f\left( \alpha  \right)} \right] + {\text{const}}.
\end{equation}
And the phase accumulation around $(k_x^0,k_y^0)$ is given by
\begin{equation}
  \Phi  = \int_0^{2\pi } {d\alpha } \,{\phi _{({k_x},{k_y})}} = 2\pi \times {\text{sign}}\left[ {\frac{{\partial \left( { h,{{\hat h}_z}} \right)}}{{\partial \left( {{k_x},{k_y}} \right)}}{|_{(k_x^0,k_y^0)}}} \right].
\end{equation}
Thus we see that, as long as the Jacobian ${\frac{{\partial \left( { h,{{\hat h}_z}} \right)}}{{\partial \left( {{k_x},{k_y}} \right)}}{|_{(k_x^0,k_y^0)}}}$ is nonvanishing, there is always a vortex at $(k_x^0,k_y^0)$ at time $t_v$. And the vorticity of this point is just the winding phase in unit of $2\pi$: $\nu = \Phi/2\pi$, giving rise to Eq.~(\ref{Eq:vorticity}). It is easy to see that the above arguments also hold for the general $n>0$ cases. Thus the vortex dynamics features a time periodicity ${t_{v,n}} = \left( {1 + 2n} \right){t_{v,0}}$.

\section{State evolution with a finite initial offset}\label{App:Finite_Mi}
For the case the initial offset $M_i$ is finite, the corresponding initial state is not polarized to the north pole. We use capital Greek letters ${\Theta _{\mathbf{k}}}$ and ${\Phi_{\mathbf{k}}}$ to parameterize the Hamiltonian vector ${\mathbf{\hat h}}\left( {\mathbf{k}} \right)$: ${\mathbf{\hat h}}\left( {\mathbf{k}} \right) = (\sin {\Theta _{\mathbf{k}}}\cos {\Phi _{\mathbf{k}}},\sin {\Theta _{\mathbf{k}}}\sin {\Phi _{\mathbf{k}}},\cos {\Theta _{\mathbf{k}}})$. The initial state is the ground state of the Hamiltonian $H_i= {\bf{h}}_i\cdot\bm{\sigma}$ (the term proportional to identity is irrelevant to the relative phase, and is thus not considered here), or written explicitly,
\begin{equation}
\left| {{\psi _{\mathbf{k}}}\left( {t = 0} \right)} \right\rangle  = \left( \begin{gathered}
  \sin \frac{{\Theta _{\mathbf{k}}^i}}{2} \hfill \\
   - \cos \frac{{\Theta _{\mathbf{k}}^i}}{2}{e^{i\Phi _{\mathbf{k}}^i}} \hfill \\
\end{gathered}  \right).
\end{equation}
After quench, the state evolves under the new Hamiltonian $H_f= {\bf{h}}_f\cdot\bm{\sigma}$, as $\left| {{\psi _{\mathbf{k}}}\left( t \right)} \right\rangle  = {e^{ - i{H_f}t}}\left| {{\psi _{\mathbf{k}}}\left( {t = 0} \right)} \right\rangle $. Now that the quench just changes the $h_z({\bf k})$ term through changing $M$, we see that $h_x({\bf k})$ and $h_y({\bf k})$ are not changed. It thus leaves the azimuthal phase ${\Phi_{\mathbf{k}}}$ invariant through the quench, i.e., we have
\begin{equation}
  {\Phi _{\mathbf{k}}^i} = {\Phi_{\mathbf{k}}^f}.
\end{equation}
We then have
\begin{equation}
  \begin{aligned}
  \left| {{\psi _{\mathbf{k}}}\left( t \right)} \right\rangle  = &\cos \left( {{h_f}t} \right)\left| {{\psi _{\mathbf{k}}}\left( {t = 0} \right)} \right\rangle  \hfill \\
  &  + i\sin \left( {{h_f}t} \right)\left( \begin{gathered}
  \sin \left( {\Theta _{\mathbf{k}}^f - \frac{{\Theta _{\mathbf{k}}^i}}{2}} \right) \hfill \\
   - \cos \left( {\Theta _{\mathbf{k}}^f - \frac{{\Theta _{\mathbf{k}}^i}}{2}} \right){e^{i\Phi _{\mathbf{k}}^i}} \hfill \\
\end{gathered}  \right). \hfill \\
\end{aligned}
\end{equation}
It is then clear that, if the conditions
\begin{equation}\label{Eq:south-pole_Mi}
\left\{ \begin{gathered}
  {\hat{\tilde {h}}_z} = \cos \left( {\Theta _{\mathbf{k}}^f - \frac{{\Theta _{\mathbf{k}}^i}}{2}} \right) = 0, \hfill \\
  \cos \left( {{h_f}t} \right) = 0, \hfill \\
\end{gathered}  \right.
\end{equation}
are satisfied, the south pole is reached. Comparing Eqs.~(\ref{Eq:condition_for_south_pole}) and (\ref{Eq:south-pole_Mi}), we see that, as long as we replace the vortex trajectory ${h_z}\left( {\mathbf{k}} \right) = 0$ for the $M_i\to\infty$ case by the first equation in Eq.~(\ref{Eq:south-pole_Mi}) [i.e., Eq.~(\ref{Eq:hz_tilde}) in the main text], the analysis of the vortex dynamics for the finite $M_i$ case takes the same steps as the one with $M_i\to\infty$.
\\

\end{document}